\documentclass{nature_print}

\def\e{\begin{equation}}
\def\f{\end{equation}}
\def\_#1{{\bf #1}}

\def\.{\cdot}

\newcommand{\p}{\varphi}
\newcommand{\wpp}{\omega_p}
\newcommand{\wc}{\omega_c}
\newcommand{\bl}{\beta_L}
\newcommand{\pea}{\varphi_{\rm ea}}

\newcommand{\wt}{\omega t}

\title{A multi-stable switchable metamaterial}

\author{P. Jung$^{1}$, S. Butz$^{1}$, M. Marthaler$^2$, M. V. Fistul$^{3,4}$, J. Lepp\"akangas$^5$, V. P. Koshelets$^{4,6}$, and A. V. Ustinov$^{1,4}$}

\begin{document}

\maketitle

\begin{affiliations}
\affil{1}{ Karlsruhe Institute of Technology, Physikalisches Institut, 76131 Karlsruhe, Germany}
\affil{2}{Karlsruhe Institute of Technology, Institut f\"ur Theoretische Festk\"orperphysik, 76131 Karlsruhe, Germany}
\affil{3}{Ruhr-Universit\"at Bochum, Theoretische Physik III, 44801 Bochum, Germany}
\affil{4}{National University of Science and Technology MISIS, Moscow 119049, Russia}
\affil{5}{Microtechnology and Nanoscience, MC2, Chalmers University of Technology, 412 96 G\"oteborg, Sweden}
\affil{6}{Kotel'nikov Institute of Radio Engineering and Electronics, Moscow 125009, Russia}
\end{affiliations}

\naturestart

\begin{abstract}

The field of metamaterial research revolves around the idea of creating artificial media that interact with light in a way unknown from naturally occurring materials.
This is commonly achieved by creating sub-wavelength lattices of electronic or plasmonic structures, so-called meta-atoms, that determine the interaction between light and metamaterial. One of the ultimate goals for these tailored media is the ability to control their properties in-situ which has led to a whole new branch of tunable and switchable metamaterials.\cite{zheludev2012,anlage2010, zheludev2011, boardman2011}
Many of the present realizations rely on introducing microelectromechanical actuators or semiconductor elements into their meta-atom structures.\cite{zheludev2011} 
Here we show that superconducting quantum interference devices (SQUIDs) can be used as fast, intrinsically switchable meta-atoms. 
We found that their intrinsic nonlinearity leads to simultaneously stable dynamic states, each of which is associated with a different value and sign of the magnetic susceptibility in the microwave domain. Moreover, we demonstrate that it is possible to switch between these states by applying a nanosecond long pulse in addition to the microwave probe signal.
Apart from potential applications such as, for example, an all-optical metamaterial switch, these results suggest that multi-stability, which is a common feature in many nonlinear systems, can be utilized to create new types of meta-atoms.
\end{abstract}
\noindent
In recent years, the idea of using SQUIDs as meta-atoms in superconducting microwave metamaterials has become increasingly appealing as the attention of the community shifts towards tunability.\cite{zheludev2012,zheludev2011} Single-junction SQUIDs are in some ways similar to split-ring resonators, one of the most common meta-atom designs, as they consist of a superconducting loop interrupted by a Josephson tunnelling barrier. Unlike most other resonant meta-atoms, however, they exhibit a unique dependence of their resonance frequency on a constant external magnetic field stemming from the nonlinear nature of the Josephson junction. This effect has been widely discussed\cite{du2006,lazarides2007,maimistov2010} and demonstrated in different waveguide geometries.\cite{jung2013,butz2013,trepanier2013} Although this truly remarkable feature is a direct effect of the Josephson nonlinearity, the SQUIDs in the cited experiments were operated at excitation powers low enough to treat the junction as a quasi-linear inductive element. With increasing power, this simple model fails as the nonlinearity starts to influence the resonant behaviour. So far, this regime has not been explored experimentally.

\noindent
The essence of this work is the dynamic multistability which occurs in  non-hysteretic, single junction (rf-)SQUIDs at intermediate power levels and has already been observed in single junctions\cite{sidiqqi2005} and investigated numerically for rf-SQUID meta-atoms.\cite{lazarides2013} It should be noted that here we deal with a purely dynamic phenomenon that is not related to the multistability known from hysteretic SQUIDs.\cite{du2008,caputo2012} For a certain choice of parameters, this multistability manifests itself as a small number of simultaneously stable states, each of which corresponds to a different value of the SQUIDs magnetic flux susceptibility $\chi_{\phi}$ at the driving frequency. More importantly, a closer analysis reveals that in many cases coexisting stable states have small positive as well as  large positive and large negative values of the real part of $\chi_\phi$. This means that, depending on the current state of the meta-atom, it can either be magnetically almost transparent, strongly enhance the local magnetic field by oscillating in phase with the drive, or even counteract the same by oscillating out of phase.

\begin{figure}
\includegraphics[width=\textwidth]{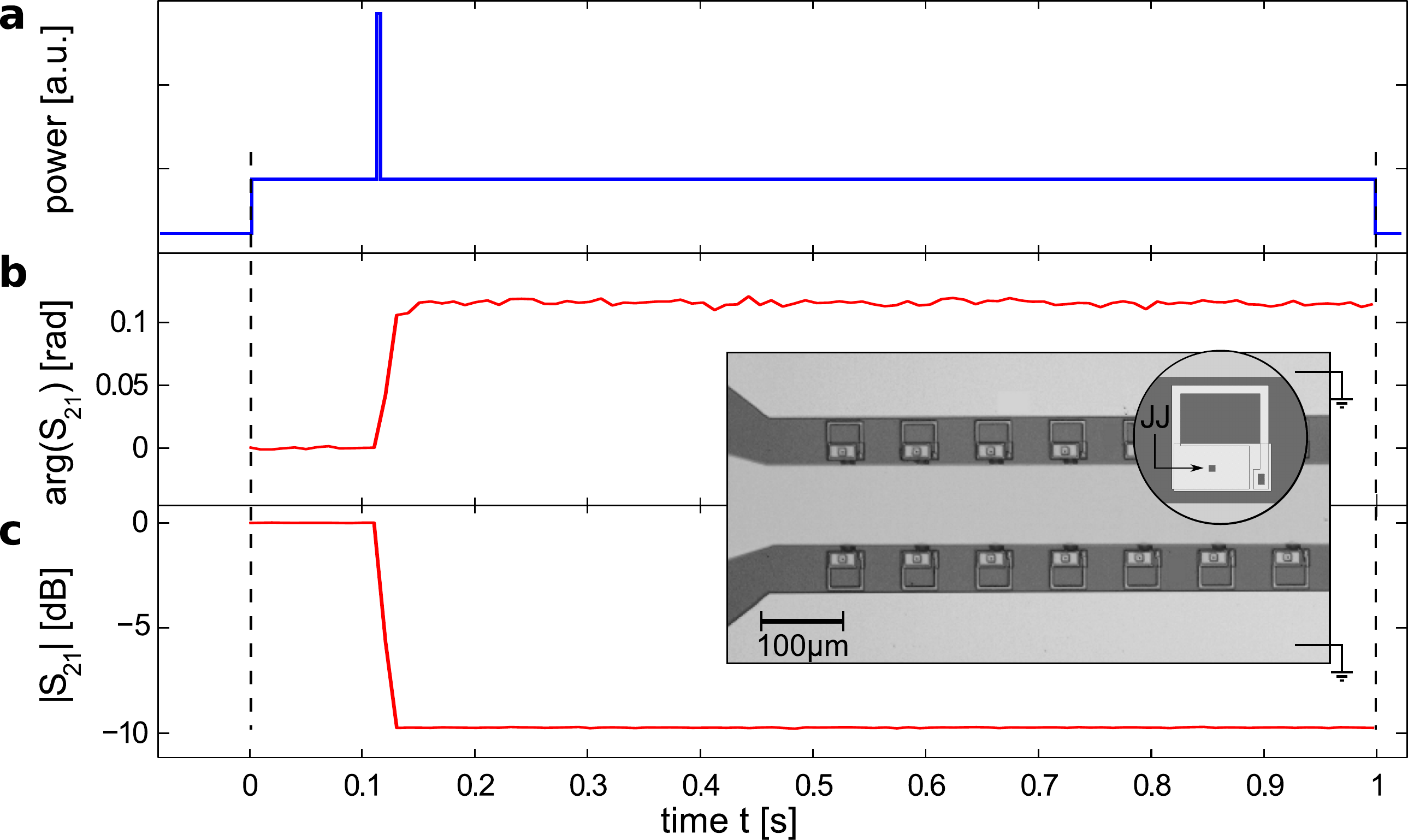}
\captionof{figure}{\textbf{a}, Sketch of the pulse sequence used to switch the one-dimensional metamaterial. The actual measurement, indicated by two vertical dashed lines, takes place after initializing the system in a small amplitude state. \textbf{b}, Transmission phase and \textbf{c} magnitude measurement of the SQUID-loaded waveguide recorded as a function of time. The steepness of the jump is limited by measurement bandwidth. The base-line has been subtracted for both curves. The pulse switches the system into another stable, dynamic state which is visible as change in transmission. The inset shows a microscope image of a section of the loaded CPW and a close-up sketch of a single SQUID. The arrow in the round inset labelled JJ marks the position of the Josephson junction. }
\label{fig1}
\end{figure}

\noindent
This is a highly desirable feature for a meta-atom provided one can reliably switch it from one into another dynamic state in-situ. Here we demonstrate this experimentally, provide an analytical description and show how they could potentially be implemented in a metamaterial device.

\multicolinterrupt{
\begin{figure}
\includegraphics[width=\textwidth]{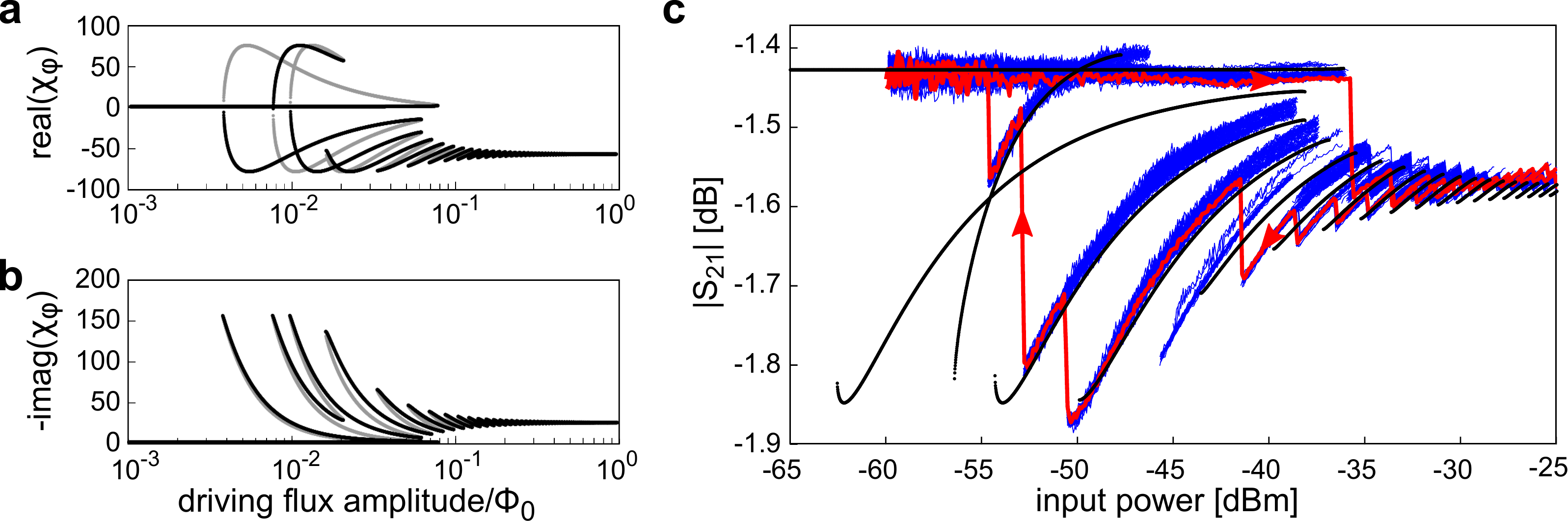}
\captionof{figure}{{\bf a}, Real and {\bf b}, imaginary part of the magnetic flux susceptibility of a single SQUID meta-atom calculated from eqns. \ref{coupledeqnx}--\ref{potrotframe}. Stable solutions are shown in black, while meta-stable solutions are in grey. {\bf c}, Calculated (black) and measured (blue and red) transmission through the sample arrangement containing only a single SQUID. The red data show a hysteresis loop from low to high power and back. The red arrows indicate the direction of the sweep. The blue data are a collection of different power sweeps of varying length and initial conditions. The over-all transmission of less than -1.4 dB is caused by the experimental setup and calibration and not related to the performance of the meta-atom.}
\label{fig2}
\end{figure}
}

\noindent
The most prominent feature associated with the effect described below, is that it is possible to switch between the states reproducibly by applying a short microwave pulse to the metamaterial. To verify this experimentally, we used a coplanar waveguide (CPW) loaded with a one-dimensional metamaterial consisting of 54 rf-SQUIDs. A row of 27 SQUIDs is located in each of the two gaps of the CPW. The coupling between the waveguide and the SQUID is designed to be primarily magnetic, which is why a change in the magnetic susceptibility seen by the waveguide can be measured as a change in its scattering characteristics. Fig. \ref{fig1} shows an exemplary transmission measurement recorded in such an experiment. Initially, the system is in a state with a susceptibility close to zero which results in little scattering and thus a large transmission through the loaded CPW. By applying the microwave pulse, the system is driven into a state with $|\chi_\phi| \gg 1$ which results in both enhanced reflection and absorption of the probe signal. Together, this causes the amount of transmitted power to drop by one order of magnitude (10 dB) for this sample. The system remains stable in the new state until it is reset.

\noindent
One noteworthy difference when comparing this system to other switchable metamaterials is that here, the switching process is rather fast because it only relies on the intrinsic properties of the nonlinear resonator. Thus the time required to switch is proportional to the decay time $\tau_d = 2 R C$ which is of the order of a few nanoseconds for our samples. Here, $R$ and $C$ are the intrinsic resistance and capacitance of the SQUID meta-atom. Experimentally we were able to verify that the switching takes place for pulses as short as $2\,$ns.

\noindent
To better understand and predict the existence of the multiple stable, dynamic states, we developed an analytical model that is able to quantitatively reproduce most of the experimental observations. We start from the basic differential equation that defines the evolution of the total gauge-invariant phase difference  $\varphi$ over the Josephson junction of the SQUID when driven externally by $\varphi_{\rm ext}$. 
\begin{equation}\label{squideqn}
\varphi + \beta_L \left [ \sin \varphi  + \frac{1}{\omega_c} \dot \varphi +  \frac{1}{\omega_p^2}\ddot \varphi \right ] = \varphi_{\rm ext},
\end{equation}
where $\beta_L = \frac{2  \pi L I_c}{\Phi_0}$, $\omega_c = \frac{2 \pi R I_c}{\Phi_0}$, $\omega_p^2 = \frac{2 \pi I_c}{C \Phi_0}$, $\Phi_0 = \frac{h}{2e}$, $I_c$ is the maximum supercurrent that can be carried by the Josephson junction, $C$ is the capacitance shunting the junction, $L$ is the inductance of the loop, and R is the damping. 
Assuming a harmonic drive $\varphi_{\rm ext} = \varphi_{\rm ea} \cos{\omega t}$, we can make an ansatz for the solution of the equation of motion: $\p = x\cos(\wt) - y \sin(\wt) $. This corresponds to all harmonic solutions that have an amplitude $A = \sqrt{x^2+y^2}$ and a phase $\delta = \arctan(y/x)$ with respect to the drive. The magnetic flux susceptibility can then be defined as the time average of the ratio between the flux in the loop and the driving flux which can be expressed as $\chi_\phi = \left < \varphi / \varphi_{\rm ext} \right >_t - 1$.
It should be noted at this point, that it is not obvious that this approach will describe the dynamics of the system sufficiently well, because it does not contain contributions from higher harmonics. A detailed treatment of why this approximation is applicable in our case is given in the methods section.
Using the mentioned ansatz, one can write down the equations of motion in the rotating wave approximation together with the stability condition:
\begin{eqnarray}
\bl\frac{\omega}{\wpp^2} \dot y& = &-\frac{\partial g}{\partial x} - \bl \frac{\omega}{\wc}y  \stackrel{!}{=} 0 \label{coupledeqnx}\\
\bl\frac{\omega}{\wpp^2} \dot x& = &\frac{\partial g}{\partial y} - \bl \frac{\omega}{\wc}x \stackrel{!}{=} 0 \label{coupledeqny}
\end{eqnarray}
Here, $g$ is the potential in the rotating frame\cite{dykman1998}:
\begin{equation}\label{potrotframe}
\begin{array}{rl}
g =&\frac{1}{2}(1-\bl \frac{\omega^2}{\wpp^2})(x^2+y^2) - \pea x -\\
&2 \bl J_0(\sqrt{x^2 + y^2}) ,\\
\end{array}
\end{equation} 
where $J_0$ is the 0th order Bessel function of the first kind. These equations yield both stable and meta-stable solutions which can be distinguished from each other through a stability analysis.

\noindent
Figures \ref{fig2}a and b show the real and imaginary part of the calculated magnetic flux susceptibility for a fixed frequency as a function of the amplitude of the drive. For a certain range of driving power, multiple stable (black) and meta-stable (grey) solutions coexist. Moreover, the simultaneously stable states can have a small positive as well as a large negative and a large positive real part of $\chi_\phi$. These results can then be used in a transmission-line approach to calculate the transmitted power through the sample as seen in Fig. \ref{fig2}c. 
Qualitatively, this effect exists for a wide range of frequencies and magnetic field values however with varying magnitude. Here, we focus on a region slightly above the LC resonance frequency $f_{\rm LC} =  ( 2 \pi \sqrt{LC} )^{-1}$ where the effect is most pronounced. 

\noindent
The fact that multi-stability only occurs in a limited range of driving power\cite{bishop2010} can be understood from the particular form of the Josephson nonlinearity $\sin \varphi$ in Eqn. \ref{squideqn}. 
For very small driving amplitudes, this term can be linearized around $\p_{\rm const}$ which in turn is related to the dc magnetic flux seen by the SQUID loop. For this reason, the SQUID is usually treated as a dc-flux tunable, quasi-linear resonator in this driving regime.
For very large driving amplitudes the nonlinear term in Eq. \ref{squideqn} is strongly suppressed. As a result, the system can be described as a dc-flux independent RLC circuit resonating at $f_{\rm LC}$ in this regime. Consequently, the only case in which the nonlinearity contributes to the dynamic behaviour of the system is in the intermediate range of driving amplitudes. Our analysis also shows that both the number of observable states and their amplitudes strongly depend on the difference between driving frequency and $f_{\rm LC}$.

\noindent
To test the analytical model and verify the existence of the dynamic states, we demonstrate two ways of measuring them: By sweeping power up and down thus going through a hysteresis loop and by exciting the system using microwave pulses. We first demonstrate the hysteresis (red curve in Fig. \ref{fig2}c) for a single SQUID meta-atom in order to obtain results that can be compared directly to the single SQUID analytic solution. At a low initial power level, the systems starts out in a transparent state (small absolute value of $\chi_\phi$) resulting in a large value of the transmission $|S_{21}|$ through the SQUID-loaded CPW. When increasing the power, it remains in this state until it is no longer stable ($\approx -37\, {\rm dBm}$ in this case). The system then settles in one of the excited states (large absolute value of $\chi_\phi$) which results in a drop in the transmitted power. Following a series of small jumps the system eventually reaches a stable level. Upon sweeping the power back, the observed transmission takes a different path by following several of the highly excited states down until they become unstable then jumping to the closest stable one. 
Although being consistent with the model, this method only reveals a small fraction of the predicted stable states. For a more complete picture one can prepare the system in each of the states using the hysteresis and then follow each path in both directions until it becomes unstable. The resulting curves are shown as blue lines in Fig. \ref{fig2}c.
One noteworthy aspect in this figure is that all but one of the highly excited states correspond to solutions with a negative real part of $\chi_\phi$ as seen in Fig. \ref{fig2}a. The only solution with a large positive real part of $\chi_\phi$ in Fig. \ref{fig2}a corresponds to the line that crosses the base-line at about $ -50\, {\rm dBm}$ driving power in Fig. \ref{fig2}c. 
Although the data agree with the theoretical prediction to a quantitative level in many aspects, there is still a slight discrepancy between the two. The most noticeable difference is that even varying the initial conditions for each sweep, we are not able to prepare the system in all states predicted by the model.

\noindent
As mentioned above, this approach can be extended from a single meta-atom to a metamaterial. In this case, however, even without any disorder, the multistability will not have such a clear signature in transmission measurements any more. This is due to the fact that the probe signal gets absorbed and reflected as it passes through the one-dimensional metamaterial so that every meta-atom sees a different effective input power. 

\begin{figure}
\includegraphics[width=\textwidth]{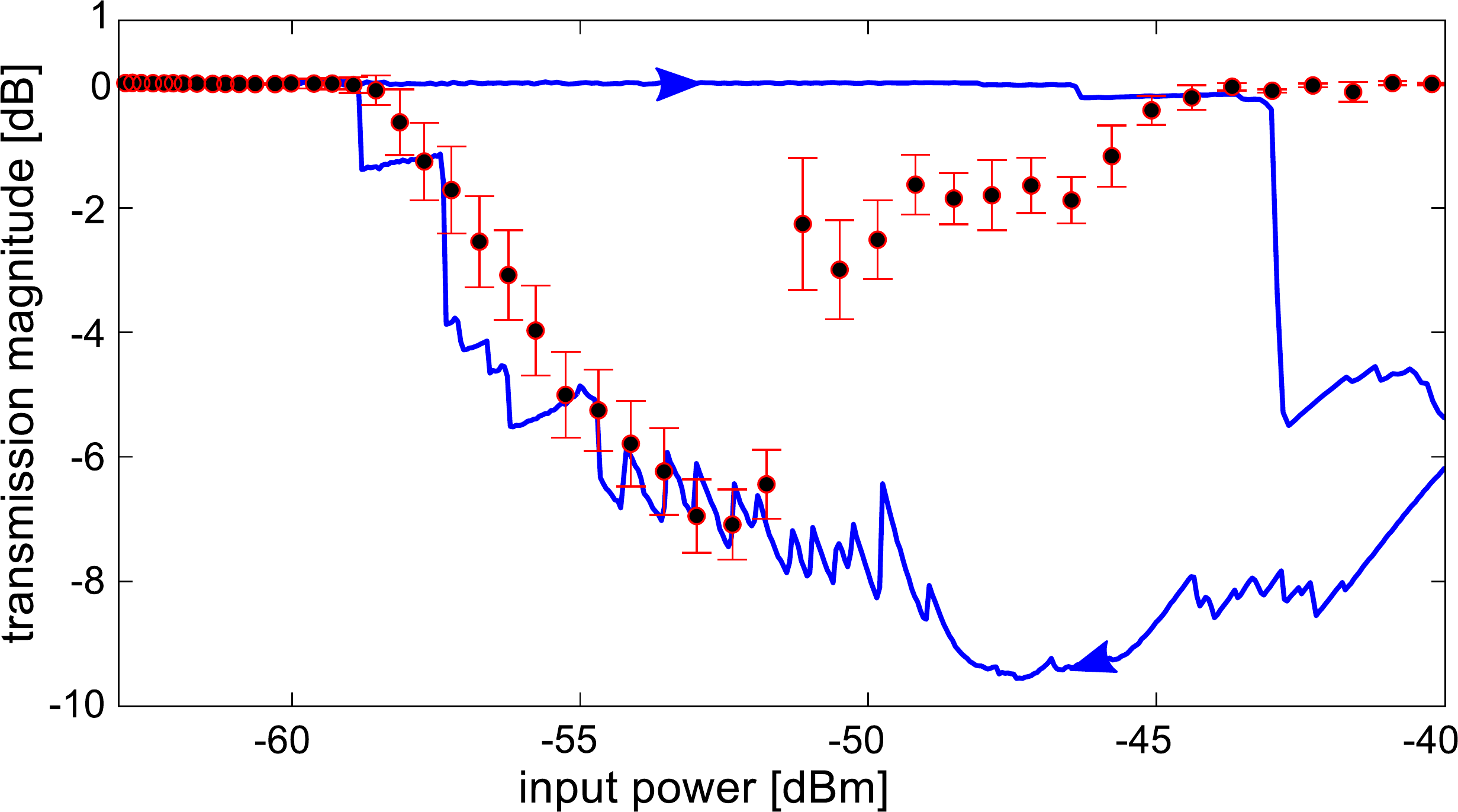}
\captionof{figure}{Hysteresis curve for a CPW loaded with 54 SQUIDs (blue). The baseline has been subtracted so it can be compared to the black dots. They show the mean magnitude of the change in $|S_{21}|$ after applying a microwave pulse as a function of the driving power. Up to a driving power of about -58\,dBm this quantity is well described by the difference between the lower and upper branch of the hysteresis. Above that threshold, the switching magnitude drops to a lower value. }
\label{fig3}
\end{figure}

\noindent
In case of the one-dimensional 54 SQUID array, one can still observe the characteristic hysteresis loop (Fig. \ref{fig3}) albeit without the distinct response of the different states. Figure \ref{fig1} shows that in this case, it is possible to decrease the transmission by up to 10 dB by applying a short microwave pulse. With the knowledge of the theory and the hysteresis curve, the pulse sequence shown in Fig. \ref{fig1}a becomes clear: The first part is necessary to reset the system to the transparent state. The intermediate power level (usually just referred to as input or driving power) moves the system to the multi-stable power region, but keeping it in the transparent branch. The pulse, finally, excites the system into a different branch of the hysteresis. The jump height is thus defined by the input power level and subject to a statistical spread. 
A more detailed overview of the switching is given in Fig. \ref{fig3}, which shows the average magnitude of the switching events as a function of the drive power. As expected, it is closely related to the difference between the up and down sweep paths of the hysteresis. Apart from the driving power level, the switching also depends on duration and amplitude of the pulse, all of which influence the number of switched SQUIDs in the array. Their statistical distribution could potentially be used to build an adjustable microwave switch.

\noindent
In conclusion, we have implemented and demonstrated a one-dimensional metamaterial consisting of switchable meta-atoms. By operating non-hysteretic, single-junction SQUIDs in an intermediate power range, we showed that they can be switched between different dynamic states corresponding to different values of their magnetic susceptibility. For individual meta-atoms, we were able to map these states directly in a transmission experiment. We developed an analytical model based on the rotating wave approximation which shows excellent agreement with the measured data.
Using an array containing 54 SQUID meta-atoms, we showed that one can switch the transmission through this  one-dimensional metamaterial by applying a nanosecond-long microwave pulse. Unlike in other switchable meta-atom implementations, we employ an intrinsic nonlinearity which makes these SQUIDs very fast-switching meta-atoms.

\begin{methodsummary}
\item[The samples.] The SQUIDs are thin-film Nb structures with an Nb-AlOx-Nb tunnel barrier. They are approximately $45\times 49\,\mu m^2$ in size with a $2\,\mu m^2$ junction area which is shunted by a large $Nb_2O_5$ capacitor. They are positioned in the gaps of a CPW on a 4x4\,mm Si chip.
The values used to model the SQUIDs are $\omega_p = 5.276 \cdot 10^{10} s^{-1} $, $\omega_c = 5.438 \cdot 10^{12} s^{-1}$, $\bl = 0.45$.

\item[Experimental setup.] All measurements were performed in liquid $^4$He at a temperature of 4.2\,K using cryogenic amplification. The sweeps were recorded with a vector network analyser. Pulses were generated using a microwave mixer and a pulse generator. 
\item[Data calibration.] The measurements were thru-calibrated at the connector closest to the sample. Additional scatterers that could not be calibrated out this way (like bond-connections) were accounted for whenever theory and experiment were compared.

\end{methodsummary}


\begin{addendum}
 \item The authors would like to acknowledge interesting and productive discussions with S. M. Anlage and N. Lazarides. This work was supported by the EU project SOLID, the Deutsche Forschungsgemeinschaft (DFG), and the State of Baden-W\"urttemberg through the DFG-Center for Functional Nanostructures (CFN) within subproject B3.5. This work was also supported in part by the Ministry of Education and Science of the Russian Federation and the Russian Foundation of Basic Research. P.J. would like to acknowledge financial support by the Helmholtz International Research School for Teratronics (HIRST).
 \item[Author Contributions] P.J. proposed and set up the experiment and wrote the paper assisted by S.B., M.M. and A.V.U. The sample fabrication was managed by V.P.K. The experiment itself was carried out by S.B. and P.J., the theory developed by M.M., J.L. and M.V.F.
 \item[Competing Interests] The authors declare that they have no
competing financial interests.
 \item[Correspondence] Correspondence and requests for materials
should be addressed to P.J.~(email: philipp.jung@kit.edu).
\end{addendum}

\natureend
\begin{methods}
\item[Validity of the RWA.]
In general the solution for the phase $\varphi$ is of the form,
\begin{equation}
 \varphi=x_0+\sum_{n=1}^{\infty}\left[x_n \cos(n\omega t)+y_n \sin(n\omega t)\right]\,.
\end{equation}
In equation \ref{potrotframe}, we only keep the resonant components $y_1=y$ and $x_1=x$ according to the rotating wave approximation.
Since the anharmonicity is relatively large, we also want
to explicitly control the validity of this approximation. If we go beyond the rotating wave approximation
not only the first mode, oscillating with frequency $\omega$ will be excited, but also higher modes.
The closer the higher modes are to the first mode, the stronger they should be excited. Therefore, we keep the 
second and third mode oscillating at $2\omega$ and $3\omega$, respectively. 

In this case, the quasi-energy Hamiltonian takes the form
\begin{eqnarray}
 g &=&\frac{1}{2}\left(1-\beta_L\frac{\omega^2}{\omega_p^2}\right)
(x_1^2+y_1^2)-\varphi_{\rm ea}x
-2\beta_L J_0(\sqrt{x_1^2+y_1^2})\cos(x_0)\\
& &+\sum_{n=2,3}\left[
 \frac{1}{2}\left(1-\beta_L\frac{n^2\omega^2}{\omega_p^2}\right)(x_n^2+y_n^2)\right.\\
 & &      \left.     -\beta_L J_n(\sqrt{x_1^2+y_1^2})\sin(x_0+\pi n/2)\left(x_n\cos(n\delta)-y_n\sin(n\delta) \right)    \right]\nonumber\\
& &+\frac{1}{\beta_L}y_0^2+x_0^2\nonumber
\end{eqnarray}
where $y_0$ is the conjugate variable of $x_0$, $J_n$ the n-th order bessel functions of the first kind and we assumed that $x_n$ and $y_n$ remain small for $n=2,3$. The resulting equations of motion for $n\geq 1$, are given by
\begin{eqnarray}
 \beta_L \frac{n\omega}{\omega_p^2}\dot{y}_n &=& -\frac{\partial g}{\partial x_n} -\beta_L\frac{n\omega}{\omega_c}y_n\,.\\
 \beta_L \frac{n\omega}{\omega_p^2}\dot{x}_n &=&  \frac{\partial g}{\partial y_n} -\beta_L\frac{n\omega}{\omega_c}x_n\,.
\end{eqnarray}
And for $n=0$ we get
\begin{eqnarray}\label{eq_Eq_o_Motion_for_the_phase}
 \frac{1}{\omega_p}\dot{y}_0 &=& -\frac{\partial g}{\partial x_0} - \frac{\omega_p}{\omega_c}y_0\,.\\
 \frac{1}{\omega_p}\dot{x}_0 &=& \frac{\partial g}{\partial y_0}\,. 
\end{eqnarray}
Equation \ref{eq_Eq_o_Motion_for_the_phase} has a stationary state ($\dot{y}_0=\dot{x}_0=0$) for $x_0=0$ independent of $x_1$ and $y_1$.
In this case we only have to consider the coupling between the resonant mode and $x_3$, $y_3$ because the first mode will not
couple to the second mode. In lowest order perturbation theory 
we calculate the solution for $x_1$ and $y_1$ as discussed above and find an order of magnitude estimate from the stationary equation of motion
\begin{eqnarray}
 0 &=& -\frac{\partial g}{\partial x_3} -\beta_L\frac{3\omega}{\omega_p}y_3\\
 0 &=&  \frac{\partial g}{\partial y_3} -\beta_L\frac{3\omega}{\omega_p}x_3
\end{eqnarray}
Solving this, we get for the amplitude of the oscillations of the third mode
\begin{equation}
A_3 = \sqrt{x_3^2+y_3^2}=\frac{\beta_L|J_3(\sqrt{x_1^2+y_1^2})|}
{\sqrt{\left(1-9\beta_L\frac{\omega^2}{\omega_p^2}\right)^2+9\beta_L^2\frac{\omega^2}{\omega_c^2}}}.
\end{equation}
On resonance we have $\beta_L\frac{\omega^2}{\omega_p^2}=1$, thus we can simplify the expression to
\begin{equation}
 A_3 \approx \frac{\beta_L}{8}|J_3(\sqrt{x_1^2+y_1^2})| = \frac{\beta_L}{8}|J_3(A_1)|.
\end{equation}
The ratio between the amplitude of the third harmonic to the amplitude at the driving frequency $A_3/A_1$ is a useful figure to quantify the validity of the RWA. We can find an upper limit for this quantity since the Bessel-Function divided by its argument can be estimated by $|J_3(A_1)|/A_1 <1/9$ and $\beta_L<1/2$. Thus $A_3/A_1 <1/144$ which means that the rotating wave approximation is well justified. \\
The frequency range of interest for the experiment at hand is determined by the low power tuning range of the resonance frequency which is bounded by $\omega_{\rm max} = \omega_p \sqrt{\frac{1+\bl}{\bl}}$ from above and $\omega_{\rm min} = \omega_p \sqrt{\frac{1-\bl}{\bl}}$ from below. In this frequency range $A_3/A_1<1/63$, which is still well within the regime of the rotating wave approximation.

\end{methods}

\end{document}